\documentclass[twocolumn,twoside,preprintnumbers,amsmath,pacs]{revtex4}
\usepackage{epsfig}
\usepackage{graphicx}
\usepackage{dcolumn}
\usepackage{bm}
\usepackage{fancyhdr}
\usepackage{pslatex}

\begin{document}

\title{{\Large Gribov ambiguities in the maximal Abelian gauge}}
\author{D. Dudal$^a$}\email{david.dudal@ugent.be}\altaffiliation{Postdoctoral fellow of the
\emph{Special Research Fund} of Ghent University.}
\author{M.A.L. Capri$^b$}
\email{marcio@dft.if.uerj.br}
\author{J.A. Gracey$^{c}$}
    \email{jag@amtp.liv.ac.uk}
\author{V.E.R. Lemes$^{b}$}\email{vitor@dft.if.uerj.br}
\author{R.F. Sobreiro$^b$}
 \email{sobreiro@dft.if.uerj.br}
\author{S.P. Sorella$^b$}
\email{sorella@uerj.br} \altaffiliation{Work supported by FAPERJ,
Funda{\c c}{\~a}o de Amparo {\`a} Pesquisa do Estado do Rio de
Janeiro, under the program {\it Cientista do Nosso Estado},
E-26/151.947/2004.}
\author{R. Thibes$^b$}
\email{thibes@dft.if.uerj.br}
\author{H. Verschelde$^a$}
 \email{henri.verschelde@ugent.be}
 \affiliation{\vskip 0.1cm $^a$
Ghent University
\\ Department of Mathematical
Physics and Astronomy \\ Krijgslaan 281-S9 \\ B-9000 Gent,
Belgium\\\\\vskip 0.1cm $^b$
 UERJ - Universidade do Estado do Rio de
Janeiro\\Rua S\~{a}o Francisco Xavier 524, 20550-013
Maracan\~{a}\\Rio de Janeiro, Brasil\\\\
\vskip 0.1cm $^c$ Theoretical Physics Division\\ Department of
Mathematical Sciences\\ University of Liverpool\\ P.O. Box 147,
Liverpool, L69 3BX, United Kingdom }

\begin{abstract}
The effects of the Gribov copies on the gluon and ghost propagators are
investigated in $SU(2)$ Euclidean Yang-Mills theory quantized in the maximal
Abelian gauge. By following Gribov's original approach, extended to the
maximal Abelian gauge, we are able to show that the diagonal component of
the gluon propagator displays the characteristic Gribov type behavior. The
off-diagonal component is found to be of the Yukawa type, with a dynamical
mass originating from the dimension two gluon condensate, which is also
taken into account. Furthermore, the off-diagonal ghost propagator exhibits
infrared enhancement. Finally, we make a comparison with available lattice
data.
\end{abstract}

\maketitle

\section{Introduction}

Nowadays, the main problem of Yang-Mills theories (YM) and consequently of
quantum chromodynamics (QCD), is to explain theoretically the confinement
phenomenon. Despite the fact that there are several approaches to treat the
issue, there is no final answer to the confinement problem. Let us briefly
point out the main ideas which will be the motivation of the present work.

\subsection{Dual superconductivity and confinement}

An appealing mechanism to explain the color confinement is the so
called dual superconductivity mechanism \cite{scon}. According to
this proposal, the low energy regime of YM would contain monopoles
as vacuum configuration. The ensuing magnetic condensation would
induce a dual Meissner effect in the chromoelectric sector. As for
ordinary superconducting media, the potential between
chromoelectric charges increases linearly with their length,
characterizing a confinement picture.

The splitting between the diagonal and off-diagonal degrees of
freedom in this approach \cite{scon,'tHooft:1981ht} indicates that
the natural way to treat the problem would be to consider
different gauge fixings for the diagonal and off-diagonal sectors
of the theory. In fact, the class of Abelian gauges
\cite{'tHooft:1981ht} shows itself to be the suitable framework to
work with. In this class of gauges not only are the diagonal and
off-diagonal sectors independently gauged, but also monopoles show
up as defects in the gauge fixing. An interesting example of an
Abelian gauge is the maximal Abelian gauge (MAG)
\cite{'tHooft:1981ht}, which will be used in this work and
discussed in the next section.

\subsection{Abelian dominance and dynamical gluon mass}

Another important ingredient of the infrared regime of QCD is the hypothesis
of Abelian dominance \cite{Ezawa:bf}. This principle states that in the
infrared limit, QCD would be described by an effective theory constructed
from only Abelian degrees of freedom. This effect has been confirmed using
lattice numerical simulations \cite{Suzuki:1989gp,Hioki:1991ai}.

Recently, it has been argued that the off-diagonal gluon might acquire a
large dynamical mass in the MAG \cite{Dudal:2004rx}, with the explicit $%
SU(2) $ value of $m\approx 2.25\Lambda _{QCD}$, due to the
condensation of the off-diagonal gluon composite operator $A_{\mu
}^{a}A_{\mu }^{a}$. This can be regarded as evidence in favour of
the Abelian dominance, since for an energy scale below this mass,
the off-diagonal gluons should decouple and the diagonal degrees
of freedom would dominate the theory. At the same level, evidence
of Abelian dominance has been advocated in the Landau gauge, due
to the condensation of dimension two operators
\cite{Capri:2005vw}. The Abelian degrees of freedom are identified
here with the diagonal gluons.

\subsection{Gribov ambiguities and the quantization of QCD}

It is a fact that YM theories are plagued by Gribov ambiguities \cite
{Gribov:1977wm,Singer:1978dk}, \textit{i.e.}, after the gauge fixing of the
model, there still remains a residual gauge symmetry spoiling a consistent
complete quantization of QCD.

The improvement of the Faddeev-Popov quantization formula
\cite{Gribov:1977wm} in the Landau gauge, yields modifications of
the propagators of the theory. The gluon propagator turns out to
be suppressed in the infrared limit, acquiring pure imaginary
poles, indicating a destabilization of the gluon excitations. On
the other hand, the ghost propagator shows itself to be more
singular than the perturbative behavior, and can be related to the
existence of long range forces. Thus, one can tacitly infer that
the Gribov problem is related to the confinement phenomenon. This
is a strong motivation to study Gribov ambiguities in the MAG.
However, concerning the Gribov problem in gauges other than the
Landau and Coulomb ones, the available information is very humble.
To our knowledge, the only available results are those obtained in
the linear covariant gauges \cite{Sobreiro:2005vn} and in the MAG
\cite{Quandt:1997rg,Bruckmann:2000xd,Capri:2005tj}. The latter one
will be discussed here.

\subsection{Lattice data}

The final motivation for the present work is related to lattice
numerical simulations, which provide a useful nonperturbative
method to treat QCD. In the last decades the lattice has been used
to extract nonperturbative effects of QCD. Mass parameters are
commonly used to fit the lattice data with relative success. In
particular, in the MAG, two mass scales have been employed to fit
the data obtained for the diagonal and off-diagonal gluon
propagators, yielding evidence in favour of the Abelian dominance
\cite {Amemiya:1998jz,Bornyakov:2003ee}. In
\cite{Bornyakov:2003ee}, the off-diagonal mass has been found to
be almost twice as big as compared to the diagonal one. Moreover,
the most suitable fit for the transverse off-diagonal gluon
propagator is found to be of the Yukawa type
\begin{equation}
D_{off}(q)=\frac{1}{q^{2}+m_{off}^{2}}\;,  \label{fit1}
\end{equation}
while for the diagonal sector, the best fit of the transverse propagator is
given by a Gribov type formula
\begin{equation}
D_{diag}(q)=\frac{q^{2}}{q^{4}+m_{diag}^{4}}\;,  \label{fit2}
\end{equation}
where
\begin{equation}
m_{off}\approx 2m_{diag}\;.  \label{Abel1}
\end{equation}
The numerical value of the off-diagonal mass is, in both works
\cite {Amemiya:1998jz,Bornyakov:2003ee}, $m_{off}\approx 1.2GeV$.
Both propagators are suppressed in the infrared limit. In
addition, a longitudinal component arises in the off-diagonal
sector from the lattice data \cite{Bornyakov:2003ee}. This
component is very well described by a Yukawa fit, (\ref{fit1}),
using the same off-diagonal mass value. In the case of the
diagonal sector, there is no longitudinal component of the
propagator.

\subsection{Analytic results}

The above considerations are sufficient to justify the study of
the MAG. Here, we shall summarize the analytic results obtained in
\cite{Capri:2005tj}, concerning the effects of the Gribov
ambiguities and of the dynamical mass generation on the
propagators of the theory, in the case of $SU(2)$ MAG.

The original Gribov approach, \cite{Gribov:1977wm}, can be
essentially repeated in the case of the MAG, where the effect of
the dynamical off-diagonal gluon mass \cite{Dudal:2004rx} can also
be taken into account. The resulting behavior of the propagators
is as follows. The transverse off-diagonal gluon propagator is of
the Yukawa type
\begin{equation}
D_{off}(q)=\frac{1}{q^{2}+m^{2}}\;,  \label{off1}
\end{equation}
where $m$ is the dynamical mass generated due to the condensation of the
off-diagonal gluon operator $A_{\mu }^{a}A_{\mu }^{a}$. The diagonal
propagator is purely transverse, being given by
\begin{equation}
D_{diag}(q)=\frac{q^{2}}{q^{4}+\gamma ^{4}}\;,  \label{diag1}
\end{equation}
where $\gamma $ is the so called Gribov parameter, with the dimension of a
mass. Further, the off-diagonal ghost propagator turns out to be enhanced in
the low energy region, according to
\begin{equation}
\lim_{q\rightarrow 0}G(q)\propto \frac{1}{q^{4}}\;.  \label{ghost1}
\end{equation}
To our knowledge this is the first result on the ghost propagator in the MAG
when the Gribov ambiguities are taken into account.

These results might provide some physical insights on the nature
of the mass parameters appearing in the lattice fits. Here, one
can easily see from (\ref {off1}) and (\ref{diag1}) that the
Gribov ambiguities are responsible for the infrared suppression of
the diagonal sector, while the dynamical mass enters only the
off-diagonal sector, making it also suppressed in the infrared.

Despite the fact that the propagators (\ref{off1}) and
(\ref{diag1}) are in qualitative agreement with the lattice data
(\ref{fit1}) and (\ref{fit2}), see \cite{Bornyakov:2003ee}, we
were not able to provide specific values for $\gamma $ and $m$ due
to the lack of a more formal framework to work with. Such a
framework is currently only at hand in the Landau gauge where a
local, renormalizable Lagrangian, which takes into account the
Gribov ambiguities, is known \cite{Zwanziger}. In
\cite{Dudal:2005na,Gracey:2005cx,Gracey:2006dr}, this Lagrangian
was used to make explicit computations. In the case of the MAG,
such a Lagrangian was recently derived \cite{Capri:2006new}, but
no explicit computations have been performed yet. Moreover, it
will allow us to investigate the behavior of the longitudinal
component of the off-diagonal gluon propagator, a feature which we
were unable to address within the approximation employed in
\cite{Capri:2005tj}.

\section{Gribov ambiguities in the MAG}

First, we shall present the MAG in the case of $SU(2)$ Yang-Mills.
Then, the Gribov ambiguities will be introduced and their main
features will be briefly discussed.

\subsection{The maximal Abelian gauge}

In order to fix the gauge differently in the diagonal and
off-diagonal sectors, we decompose  the $SU(2)$ gluon field
according to \cite{Fazio:2001rm}
\begin{equation}
\mathcal{A}_{\mu }=A_{\mu }^{a}T^{a}+A_{\mu }T^{3}\;,  \label{dec1}
\end{equation}
where $T^{a}$, $a=1,2$, stands for the off-diagonal generators of $SU(2)$,
while $T^{3}$ denotes the diagonal one. In the same way the field strength
decomposes as
\begin{equation}
\mathcal{F}_{\mu \nu }=F_{\mu \nu }^{a}T^{a}+F_{\mu \nu }T^{3}\;,
\label{dec2}
\end{equation}
so that, for the YM action we get
\begin{equation}
S_{YM}=\frac{1}{4}\int {d^{4}x}\left( F_{\mu \nu }^{a}F_{\mu \nu
}^{a}+F_{\mu \nu }F_{\mu \nu }\right) \;.  \label{ym1}
\end{equation}
Explicitly, for the components of the field strength we have
\begin{eqnarray}
F_{\mu \nu }^{a} &=&D_{\mu }^{ab}A_{\nu }^{b}-D_{\nu }^{ab}A_{\mu }^{b}\;,
\nonumber \\
F_{\mu \nu } &=&\partial _{\mu }{A}_{\nu }-\partial _{\nu }{A}_{\mu }\;,
\label{str1}
\end{eqnarray}
where the covariant derivative $D_{\mu }^{ab}$ is defined with respect the
diagonal gluon
\begin{equation}
D_{\mu }^{ab}=\delta ^{ab}\partial _{\mu }-g\epsilon ^{ab}A_{\mu }\;.
\label{cov1}
\end{equation}
The action (\ref{ym1}) is invariant under the gauge transformations
\begin{eqnarray}
\delta {A}_{\mu }^{a} &=&-D_{\mu }^{ab}\omega ^{b}-g\epsilon ^{ab}A_{\mu
}^{b}\omega \;,  \nonumber \\
\delta {A}_{\mu } &=&-\partial _{\mu }\omega -g\epsilon ^{ab}A_{\mu
}^{a}\omega ^{b}\;,  \label{gauge1}
\end{eqnarray}
where $\{\omega ^{a},\omega \}$ are the infinitesimal gauge parameters. The
presence of the gauge freedom requires a constraint for a consistent
perturbative quantization. The MAG gauge fixing condition is attained by
requiring that
\begin{equation}
D_{\mu }^{ab}A_{\mu }^{b}=0\;.  \label{mag1}
\end{equation}
The residual local $U(1)$ gauge symmetry present in the diagonal
sector is fixed by means of the Landau gauge
\begin{equation}
\partial _{\mu }{A}_{\mu }=0\;.  \label{mag2}
\end{equation}
According to \cite{Capri:2005tj}, the gauge fixing conditions (\ref{mag1}), (%
\ref{mag2}) yield the following partition function
\begin{equation}
\mathcal{Z}=\int {D}A{^{a}D}A\left( \det {\mathcal{M}^{ab}}\right) \delta
(D_{\mu }^{ab}A_{\mu }^{b})\delta (\partial _{\mu }{A}_{\mu })e^{-S_{YM}}\;,
\label{path1}
\end{equation}
where $\mathcal{M}^{ab}$ is the off-diagonal Hermitian Faddeev-Popov ghost
operator
\begin{equation}
\mathcal{M}^{ab}=-D_{\mu }^{ac}D_{\mu }^{cb}-g^{2}\epsilon ^{ac}\epsilon
^{bd}A_{\mu }^{c}A_{\mu }^{d}\;.  \label{ghost2}
\end{equation}

\subsection{The Gribov problem in the MAG}

The existence of a residual gauge symmetry in the path integral (\ref{path1}%
) is recognized as the well known Gribov problem \cite{Gribov:1977wm}. It
consists in the existence of equivalent gauge field configurations, $\{%
\widetilde{A}_{\mu }^{a},\widetilde{A}_{\mu }\}$, obeying the gauge
conditions (\ref{mag1}) and (\ref{mag2}), namely
\begin{eqnarray}
D_{\mu }^{ab}(\widetilde{A})\widetilde{A}_{\mu }^{b} &=&0\;,  \nonumber \\
\partial _{\mu }\widetilde{A}_{\mu } &=&0\;.  \label{gc}
\end{eqnarray}
At the infinitesimal level, conditions (\ref{gc}) yield

\begin{eqnarray}
\mathcal{M}^{ab}\omega ^{b} &=&0\;,  \label{copies1A} \\
-\partial ^{2}\omega -g\epsilon ^{ab}\partial _{\mu }(A_{\mu }^{a}\omega
^{b}) &=&0\;,  \label{copies1}
\end{eqnarray}
implying the existence of zero modes for the Faddeev-Popov operator $%
\mathcal{M}^{ab}$. Thus, the Yang-Mills measure in the partition function (%
\ref{path1}) is ill-defined. Also, from eq.(\ref{copies1}), one
observes that the diagonal parameter $\omega $ is completely
determined once eq.(\ref {copies1A}) has been solved for $\omega
^{a}$, according to

\begin{equation}
\omega =-g\epsilon ^{ab}\frac{1}{\partial ^{2}}\partial _{\mu }(A_{\mu
}^{a}\omega ^{b})\;.  \label{ngc}
\end{equation}
As expected, this observation allows us to prove that the diagonal
ghosts decouple, and can be in fact integrated out in the
partition function (\ref{path1}). Thus, one can focus only on the
zero modes of the off-diagonal Faddeev-Popov operator,
eq.(\ref{copies1A}).

\subsection{Facing the Gribov copies}

According to \cite{Capri:2005tj}, the existence of the Gribov
copies in the MAG can be faced along the lines outlined by Gribov
in the case of the Landau and Coulomb gauges
\cite{Gribov:1977wm}, where  the domain of
integration in the Feynman path integral is restricted to a smaller region $%
\Omega $, known as the Gribov region.

In the case of the MAG, the region $\Omega $ is identified with the set of
field configurations obeying the gauge conditions (\ref{mag1}) and (\ref
{mag2}), and for which the Faddeev-Popov operator, eq.(\ref{ghost2}), is
strictly positive, namely
\begin{equation}
\Omega \equiv \left\{ \mathcal{A}_{\mu }\;\Big|\;D_{\mu }^{ab}A_{\mu
}^{b}=0\;,\;\partial _{\mu }{A}_{\mu }=0\;,\;\mathcal{M}^{ab}>0\right\} \;.
\label{reg1}
\end{equation}
The boundary $\partial \Omega $ of the region $\Omega $ is known
as the Gribov horizon. In the MAG, the restriction of the domain
of integration in the path integral to the region $\Omega $ is
supported by the fact that for a field
configuration belonging to $\Omega $ and lying near the Gribov horizon $%
\partial \Omega $, there is an equivalent configuration located on the other
side of the horizon $\partial \Omega $, outside of the Gribov
region $\Omega $. This result is a generalization to the MAG of
Gribov's original statement in the Landau gauge. The complete
proof can be found in \cite {Capri:2005tj}. The restriction to the
region $\Omega $ is achieved by modifying the partition function
(\ref{path1}) in such a way that
\begin{equation}
\mathcal{Z}=\int DA{^{a}}DA\left( \det {\mathcal{M}^{ab}}\right) \delta
(D_{\mu }^{ab}A_{\mu }^{b})\delta (\partial _{\mu }{A}_{\mu })e^{-S_{YM}}%
\mathcal{V}(\Omega )\;,  \label{path2}
\end{equation}
where the functional $\mathcal{V}(\Omega )$ implements the restriction to $%
\Omega $ in field space.

The functional $\mathcal{V}(\Omega )$ can be constructed
recursively by means of a no-pole condition on the off-diagonal
ghost propagator, which is nothing else but the inverse of the
operator $\mathcal{M}^{ab}$. In fact, from the definition of the
region $\Omega $, it follows that the inverse of the Faddeev-Popov
operator, $(\mathcal{M}^{ab})^{-1}$, see \cite
{Gribov:1977wm,Capri:2005tj}, has to be positive and without
singularities, except for those configurations which are located
on the boundary $\partial \Omega ,$ where $\mathcal{M}^{ab}$
vanishes. Moving to momentum space, it can be shown that the Green
function $\mathcal{G}(k)=\left\langle k\left|
\mathcal{M}^{-1}\right| k\right\rangle $ has no poles at
nonvanishing $k^{2}$, except for a singularity at $k^{2}=0$,
corresponding in fact to the boundary $\partial \Omega $.
According to the no-pole prescription, the first nontrivial term
for the factor $\mathcal{V}(\Omega )$ if found to be \cite
{Capri:2005tj}

\begin{equation}
\mathcal{V}(\Omega )=\exp \left\{ -\frac{\gamma ^{4}}{2}\int \frac{d^{4}q}{%
(2\pi )^{4}}\frac{A_{\mu }(q)A_{\mu }(-q)}{q^{4}}\right\} \;,  \label{rest1}
\end{equation}
where $\gamma $ is the Gribov parameter. It is not a free
parameter, being determined by the gap equation
\begin{equation}
\frac{3}{4}g^{2}\int \frac{d^{4}q}{(2\pi )^{4}}\frac{1}{q^{4}+\gamma ^{4}}%
=1\;.  \label{gap1}
\end{equation}
The gap equation (\ref{gap1}), together with the path integral (\ref{path2})
ensures the correct truncation of the integration domain up to the Gribov
horizon.

\subsection{Propagators I}

The restriction of the domain of integration, eq.(\ref{path2}),
has far reaching consequences on the behavior of the propagators.
Looking at the form of the functional (\ref{rest1}), one can
easily see that it affects only the
diagonal tree level propagator. In fact, the diagonal propagator is given by (%
\ref{diag1}), which is infrared suppressed. We notice the appearance of
imaginary poles in this propagator, indicating that the diagonal gluon does
not belong to the physical spectrum of the model. This behavior is
consistent with the confining character of the theory.

Concerning the off-diagonal gluon propagator, it is left
unmodified at the tree level, coinciding with the transverse
perturbative propagator
\begin{equation}
D_{off}^{pert}=\frac{1}{q^{2}}\;.  \label{off2}
\end{equation}
There is no longitudinal off-diagonal component at the tree level.

The off-diagonal ghost propagator shows itself to be more singular
in the infrared region, as can be inferred from (\ref{ghost1}). We
remark that the gap equation (\ref{gap1}) is an essential
ingredient for this behavior. Notice also that this enhancement is
a manifestation of the Gribov horizon, where the ghost propagator
is highly singular.

\section{Dynamical mass in the MAG}

In \cite{Capri:2005tj}, the Gribov issue was also studied in the
presence of the off-diagonal dynamical gluon mass. In \cite
{Dudal:2004rx}, the condensation of the operator $A_{\mu
}^{a}A_{\mu }^{a}$ was analysed in detail in the MAG, by
evaluating the effective potential for this operator at one loop
order. It was shown that this potential develops a nontrivial
minimum, which lowers the vacuum energy, favoring thus the
formation of the condensate $\left\langle A_{\mu }^{a}A_{\mu
}^{a}\right\rangle $, which results in an effective dynamical
off-diagonal gluon mass. Following \cite{Dudal:2004rx}, the
dynamical mass generation can be described by adding to the
Yang-Mills action the following term
\begin{equation}
S_{m}=\frac{1}{2\zeta {g}^{2}}\int {d^{4}x}\left[ \sigma ^{2}+g\sigma {A}%
_{\mu }^{a}A_{\mu }^{a}+\frac{g^{2}}{4}\left( A_{\mu }^{a}A_{\mu
}^{a}\right) ^{2}\right] \;,  \label{mass1}
\end{equation}
where $\sigma $ is a Hubbard-Stratanovich auxiliary field coupled to the
composite operator $A_{\mu }^{a}A_{\mu }^{a}$. As shown in \cite
{Dudal:2004rx}, the field $\sigma $ develops a nonvanishing vacuum
expectation value, $\left\langle \sigma \right\rangle \neq 0$, which is
related to the gluon condensate $\left\langle A_{\mu }^{a}A_{\mu
}^{a}\right\rangle $ through the relation \cite{Dudal:2004rx}
\begin{equation}
\left\langle \sigma \right\rangle =-\frac{g}{2}\left\langle A_{\mu
}^{a}A_{\mu }^{a}\right\rangle \;.  \label{cond1}
\end{equation}
The parameter $\zeta $ is needed to account for the vacuum divergences
present in the Green function $\left\langle A^{2}(x)A^{2}(y)\right\rangle $.
At one loop order, for the dynamical gluon mass $m$ one finds \cite
{Dudal:2004rx}
\begin{equation}
m^{2}=\frac{\left\langle A_{\mu }^{a}A_{\mu }^{a}\right\rangle }{g\zeta }%
\approx \left( 2.25\Lambda _{QCD}\right) ^{2}\;,  \label{mass2}
\end{equation}
Remarkably, the effective action
\begin{equation}
S=S_{YM}+S_{m}\;,  \label{action1}
\end{equation}
turns out to be gauge invariant, provided the auxiliary field $\sigma $
transforms as
\begin{equation}
\delta \sigma =g{A}_{\mu }^{a}D_{\mu }^{ab}\omega ^{b}\;.  \label{gauge2}
\end{equation}
As a consequence,  the action (\ref{action1}) is still plagued by
the Gribov ambiguities. In order to cure this pathology, one can
perform the same procedure as was performed for the massless case.
According to \cite{Capri:2005tj}, both the functional
(\ref{rest1}) and the gap equation (\ref{gap1}) remain the same.

\subsection{Propagators II}

Since the functional (\ref{rest1}) depends only on the diagonal gluon field,
the corresponding propagator is not affected, at the tree level, by the
dynamical mass. Thus, the diagonal gluon propagator is given by (\ref{diag1}%
).

The off-diagonal gluon propagator is,  however, affected by the dynamical
mass. A simple computation leads to the expression (\ref{off1}), exhibiting
infrared suppression due to the dynamical mass. .

The ghost propagator, also not affected by the gluon mass, shows
the typical infrared enhancement (\ref{ghost1}). Again, the gap
equation (\ref{gap1}) is fundamental for this result.

\section{Conclusions and more}

In this work we have reviewed the influence of the Gribov ambiguities and of
the dynamical gluon mass on the propagators of $SU(2)$ Yang-Mills theories
in four-dimensional Euclidean space time, quantized in the maximal Abelian
gauge. The diagonal as well as the off-diagonal gluon propagators are
infrared suppressed. The diagonal sector displays a Gribov type behavior,
due to the presence of the Gribov parameter, see (\ref{diag1}).  The
off-diagonal gluon propagator has a Yukawa type behavior, see (\ref{off1}).

The off-diagonal ghost propagator was computed in the infrared limit. The
result (\ref{ghost1}) shows that it is enhanced in the low energy region.
To some extent, this enhancement signals the influence of field
configurations located near the boundary $\partial \Omega $ of the Gribov
region, where the ghost propagator is highly singular \cite
{Gribov:1977wm,Maas:2005qt}.

We recall here that, for the gluon propagators, our results are in
qualitative agreement with the available lattice data
\cite{Bornyakov:2003ee}, see (\ref{fit1}) and (\ref{fit2}).
Unfortunately, till now, no data for the ghost propagator are
available.

\section*{Acknowledgements}

R.~F.~Sobreiro would like to thank the organizers of the IRQCD in Rio for
the kind invitation for this talk. The Conselho Nacional de Desenvolvimento
Cient\'{i}fico e Tecnol\'{o}gico (CNPq-Brazil), the FAPERJ, Funda{\c{c}}{%
\~{a}}o de Amparo {\`{a}} Pesquisa do Estado do Rio de Janeiro, the SR2-UERJ
and the Coordena{\c{c}}{\~{a}}o de Aperfei{\c{c}}oamento de Pessoal de N{%
\'{i}}vel Superior (CAPES) are gratefully acknowledged for financial
support. D. Dudal acknowledges support from the Special Research Fund of
Ghent University. This work was supported by FAPERJ, under the program
Cientista do Nosso Estado, No. E-26/151.947/2004.





\end{document}